\documentclass[referee]{aa}
\usepackage{graphics}
\usepackage{epsfig}

\def\msol{M_\odot}

\def\te{T_{\rm eff}}

\def\simgr{\,\hbox{\hbox{$ > $}\kern -0.8em \lower 1.0ex\hbox{$\sim$}}\,}
\def\simle{\,\hbox{\hbox{$ < $}\kern -0.8em \lower 1.0ex\hbox{$\sim$}}\,}
\def\beq{\begin{equation}}
\def\eeq{\end{equation}}

\def\aj{AJ}                   
\def\apj{ApJ}                 
\def\apjl{ApJ}                
\def\apjs{ApJS}               
\def\ao{Appl.Optics}          
\def\aap{A\&A}                
\def\mnras{MNRAS}             
\def\ActaA{{ Acta Astron.}}
\def\vol#1{ #1}

\def\te{t_{\mathrm {E}}}

\def\at{A_{\mathrm{T}}}
\def\ut{u_{\mathrm{T}}}
\def\umin{{u_{\mathrm {min}}}}
\def\re{R_{\mathrm {E}}}
\def\fs{F_{\mathrm {s}}}
\def\fo{F_{\mathrm {o}}}
\def\aobs{{A_{{\mathrm {obs}}}}}
\def\bbar{\bar B}
\def\atrue{{A}}
\def\umintrue{{u_{{\mathrm {min}}}}}
\def\uminobs{{u_{{\mathrm {min,obs}}}}}
\def\tetrue{{t_{{\mathrm {E}}}}}
\def\utrue{{u}}
\def\amaxobs{{A_{{\mathrm {max,obs}}}}}

\def\tdobs{{t_{{\mathrm {d,obs}}}}}
\def\td{{t_{{\mathrm {d}}}}}
\def\teobs{{t_{{\mathrm {E,obs}}}}}

\def\fb{f}

\def\({\left(}
\def\){\right)}
\def\<{\left<}
\def\>{\right>}

\def\tauobs{\tau_{{\mathrm {obs}}}}
\def\tautrue{\tau}

\begin{document}

\title{Source blending effects on microlensing time-histograms and optical depth determination.}

\author{Yann Alibert \inst{1,2} \and Gilles Chabrier \inst{2} \and G\'erard Massacrier \inst{2}}

\institute{
$^1$ Physikalisches Institut, University of Bern, Sidlerstrasse 5, CH-3012 Bern, Switzerland \\
$^2$ C.R.A.L (UMR 5574 CNRS),
 Ecole Normale Sup\'erieure, 69364 Lyon Cedex 07, France\\ email: yann.alibert@phim.unibe.ch, chabrier@ens-lyon.fr, gmassacr@ens-lyon.fr
}

\offprints{Y. Alibert \email{yann.alibert@phim.unibe.ch}}

\date{Received 18/05/05 - in original form 22/03/01 - Accepted 28/06/05}

\abstract{
Source blending in microlensing experiments is known to modify the Einstein time of the observed events.
In this paper, we have conducted Monte-Carlo calculations, using the analytical relationships derived by Han (1999)  
to quantify the effect of blending on the observed event time distribution and optical depth. We show that short-time events are affected 
significantly by source blending and that, for moderately blended sources, the optical depth $\tau$ is globally overestimated,
 because of an underestimation of the exposure. For high blending situations, on the opposite, blending leads to an {\it under}estimation of the optical depth. Our results are in agreement with the most recent optical depth determinations toward the Galactic Center of the 
MACHO collaboration (Popowski et al. 2004) and the OGLE-II collaboration (Sumi et al. 2005)
that use clump giants (less affected by the blending effect) as sources.
The blending-corrected, lower optical depth toward the Galactic Bulge is now in good agreement with the value inferred from 
galactic models, reconciling theoretical and observational determinations.
}

\titlerunning{Source blending effects}
\authorrunning{Alibert Y., et al.} 
\maketitle

\section{Introduction}

Microlensing is one of the most efficient ways to measure the amount of (sub)stellar matter, dark or luminous, in our Galaxy.
By measuring the amplification
of the flux of a source star, one can derive statistically the mass, distance and transverse velocity of the deflector.
When a deflector passes near the line of sight between the observer and the source,
 the flux of the source is amplified by a factor (Paczy\'nski 1986)  
$$
A(u) = { u^2 + 2  \over  u \sqrt{u^2+4}}  \,\,\,\mbox{where}\,\,\, u(t)=\sqrt{{\umin}^2 + {\left( t \over \te \right)}^2}
$$
if the origin of time is taken at the moment of maximum amplification.
The light curve  depends on two parameters only, $\umin$ and $\te$, where $\umin$ is the minimum impact parameter 
(the minimum distance to the line 
of sight,
normalized to the Einstein radius of the lens $\re$), and $\te$ is the Einstein time, i.e. the time required to cross the Einstein radius.

The Einstein radius is given by the lens mass $M$, the observer-source  distance  $L$,
and the ratio $x$ of the lens-observer distance to $L$ :
$$
\re^2 = {4 G M \over c^2} L x (1-x)
$$

When a microlensing event occurs, a theoretical light curve is ajusted to give $\umin$ and $\te$: $\umin$ is related to the maximum amplification
by $\umin = f(A_{{\mathrm {max}}})$, where the function $f$ is defined by $f(x) \equiv \sqrt{{2 x \over \sqrt{x^2-1}}-2}$.
The Einstein time is obtained from the event duration $\td$ during which $A > \at$ (where $\at$ is the threshold amplification,
usually taken as $3/\sqrt{5}$, which corresponds to $\ut = 1$) by:
$$
\te = { \td \bigm/ \sqrt{{\ut}^2-{\umin}^2}}
$$

The observed optical depth is estimated from the Einstein times of the events as (Alcock et al. 1997):  
$$
\tau_{{\mathrm {obs}}} = { \pi \over 4 E } \sum_i { {\te}_i \bigm/ \varepsilon({\te}_i ) }, 
$$
where $E$ is the exposure, i.e. the number of observed sources
times the duration of the experiment (in stars$\times$years), and $\varepsilon$ is the detection efficiency.
This optical depth can be compared with the one obtained from galactic models using
\beq
  \tau  = \int {4 \pi G \over c^2} x(1-x) L^2 \rho(xL) dx
\label{eqtau}
\eeq
for a constant observer-source distance, where $\rho$ is the galactic mass density distribution.

A long standing unsolved, major problem concerning the microlensing
experiments towards the Galactic bulge is the significant discrepancy between the optical depths derived from microlensing experiments and the one
calculated from theoretical models, as identified originally by Kiraga \& Paczy\'nski (1994).       
Using Difference Image Analysis, and using main sequence stars as microlensing sources,
 the MACHO project (Alcock et al. 2000a)  
 deduced an optical depth
$\tau_{-6} = 2.91 ^{+0.47}_{-0.45}$ at $2 \sigma$ with $99$ observed events, where $\tau_{-6} \equiv \tau / 10^{-6}$. The OGLE collaboration derived
$\tau_{-6} = 3.3 \pm 2.4$ at $2 \sigma$ with $9$ observed events
(Udalski et al. 1994). More recently, the MOA collaboration has determined an optical depth 
$\tau_{-6} = 2.59^{+0.84}_{-0.64}$ (Sumi et al. 2003).  

On the other hand, all Galactic models predict a typical optical depth $\tau\simeq 1-1.2 \times 10^{-6}$ (see for example
Peale 1998,  
Bissantz et al. 1997).   
 As shown by Peale (1998),    
one can reproduce the observed optical depth only if the bulge mass is equal to $M_b=3.3 \times 10^{10} \msol$,
whereas gas dynamics calculations based on DIRBE observations to derive the gravitational potential of our Galaxy 
yield a maximum mass for the bulge $M_{b_{max}} \simeq 2 \times 10^{10} \msol$
(Bissantz et al. 1997,   
 Englmaier et al. 1999, Sevenster et al. 1999).
More recently, Han \& Gould (2003) have derived an optical depth $\tau_{-6} = 0.98$, for events due to bulge
stars, and $\tau_{-6} = 1.63$ when including the events due to the disk.

Using brighter stars (in particular red clump giants), the MACHO and the OGLE-II
collaborations have derived recently new optical depths determinations, namely $\tau_{-6} = 2.17 ^{+0.47}_{-0.38}$ for $62$ events 
for the MACHO group (Popowski et al. 2004), and $\tau_{-6} = 1.96 ^{+0.41}_{-0.34}$ for 33 events for the 
OGLE-II group (Sumi et al. 2005). These values, closer to the theoretical expectations,
are significantly lower than the aforementioned ones based on main sequence star sources. This raises two important questions:
1) what is the origin of the systematic discrepancy between optical depth measured using faint (main sequence)
and bright (clump giant) stars, and 2) which one should be compared with theoretical calculations.
We demonstrate here that this difference is, for a large part, due to blending effects, and that the
optical depth measured using main sequence stars is systematically biased.

Microlensing experiments are subject to different biases like source blending (the observed lensed source
is blended with other unresolved stars), lens blending (when the lens is a luminous star, its own luminosity
may affect the observed light curve), and amplification bias (the amplified source is a star below the detection limit
close to another star above the detection limit).
Lens blending has been adressed by Han (1998);   
the effect is found to result in an underestimation of the real optical depth by $\simeq 10\%$.
Amplification bias has been studied also by Han (1997):  
its effect 
is an overestimation  of the optical depth by a factor $\simeq 1.7 ^{+ 0.7} _{-0.4}$, depending on the size of the seeing disk. 
Alard (1997)     
studied source blending 
by taking it into account directly in Eq. \ref{eqtau}. The effect on the optical depth is an overestimation of $\sim 15 \%$, but
the effect on $\te$-histograms was not quantified.
 In this paper, we study in detail the afore-mentioned first kind of blending, i.e. the amplified source is a star
 above the detection limit, blended with a star above the detection limit.

In the case of source blending, an observed source is the superposition of different source stars, all above the detection limit.
When an event occurs, only one of the stars composing the observed source will be amplified, in general \footnote{the probability that more than one source
is amplified is negligible.}. The observed flux is no longer the baseline flux multiplied by the amplification,
but instead, if one notes $\fs$ the flux of the amplified source and $\fo$ the flux of the other unresolved sources,
$$\aobs= {A(u(t)) \fs + \fo \over \fs + \fo}$$ It is very difficult to account
correctly for this effect for, as shown by Han (1999),     
there is an almost perfect degeneracy between unblended and blended light curves. Moreover, it is very difficult to
determine the amount of blending  by photometrical means (see Wo\'zniak \& Paczy\'nski 1997).   
 Nevertheless, using
 astrometric shift during a microlensing event,
Goldberg and Wo\'zniak (1998)    
have shown that about half of the sources observed by the OGLE collaboration suffer some amount of blending.

To quantify the effect of source blending on the duration
of a microlensing event and on the inferred optical depth, 
we first outline in \S$2$ the relations between $\te$ and $\umin$ with
and without blending. In \S$3$, we derive the distribution function to be used for blending. The consequences on the Einstein time histograms 
and on the optical depth are examined in \S$4$ and  5, respectively. 
Section $6$ is devoted to the conclusions.

\section{Analytical relationships}

We will follow the previous calculations by Han (1999)     
with slightly different notations.

When a star suffers blending, the total observed flux is the sum of the flux from the source $\fs$ and the flux from the
other unresolved stars $\fo$. We define the dimensionless parameters $B$ as $B \equiv \fs / (\fs+\fo) $ and $\bbar \equiv 1 - B$.

The observed amplification of a microlensing event thus reads:
$$
\atrue (t) = B {{ \utrue ^2 + 2 } \over { \utrue \sqrt{\utrue^2+4}} }+ \bbar
$$
Using the observed light curve, one can derive some quantities, such as the impact parameter, the duration of the event.
We will label $'{{\mathrm{obs}}}'$ the quantities derived when one supposes that $B=1$ (i.e. when one ignores that
the source is blended). The quantities corrected from blending will be called 'true',
and will have no index.

The  observed impact parameter $\uminobs$ is 
\begin{eqnarray}
\uminobs  & = & f \left( \amaxobs \right) \nonumber \\
          & = & f \left( B {{ \umintrue^2 + 2 } \over { \umintrue \sqrt{\umintrue^2+4}}} + \bbar \right) \nonumber
\end{eqnarray}

\noindent so that the duration of the event
$$
\tdobs^2= \tetrue^2 \left( f^2 \left({\at - \bbar    \over  B } \right) - \umintrue^2 \right)
$$

\noindent is underestimated.

We can then relate
 the observed Einstein time to the true Einstein time, the true impact parameter and
the amount of blending by :
\beq
\teobs= \tetrue { \sqrt{ f^2 \left({\at - \bbar    \over  B } \right) - \umintrue^2} \biggm/ \sqrt{ {\ut}^2 - {\uminobs}^2  }}
\label{Fteobs}
\eeq

These relations allow to relate the 'observed' and 'real'  quantities, for a known amount of blending.
As shown by Han (1999),    
the curves characterized by $(\tetrue,\umintrue,B)$ and the corresponding
$(\teobs,\uminobs,B=1)$ lie very close to each other, and the difference is smaller than the observational errors.

\section{Blending Distribution Function}

In order to estimate the effect of blending, one needs to know the distribution function (DF) of the blending factor $B$. 
We first need to estimate the number of stars whose projection lies inside a small surface $\Delta S$.
For that purpose, we suppose that the projected
density of stars is constant and that the stars are distributed randomly. 
By doing so, we neglect all systematic effects in the spatial distribution
of stars. Note that this approximation is used by the MACHO group to compute
their detection efficiency toward the LMC (Alcock et al. 2001).     
Then, the probability that there are $n$ stars inside $\Delta S$ is given by 
the Poisson law:
$
\wp(n) = { {\bar n}^n e^{- \bar n} / n ! }
$, where $\bar n$ is the mean number of stars lying inside $\Delta S$.

The second step is to calculate the probability that there are $n$ stars inside the seeing disk, of surface $\Delta S$, of
an observed source. This probability $P(n)$ is the probability that there are $n$ stars inside $\Delta S$,
{\it knowing that there is at least one star inside $\Delta S$}. We thus have
$$
P(n) \equiv \wp\( n | n \geq 1\) = { \wp (n) \over \sum_{i=1}^{\infty} \wp(i) }  =
 { {\bar n}^n e^{- \bar n} \over n ! \( 1 - e^{- \bar n} \)}
$$
The only free parameter in this law \footnote{In the following, we will refer to $P$
as the Poisson law, although this is not exactly the standard Poisson law because of the different normalization factor} is $\bar n$, the mean number of stars inside a seeing disk.
This parameter depends on the size of the seeing disk, and we will usually
take $\bar n \simeq 1.257$, which ensures that 
half of the observed sources are unblended (i.e. $P(1) = 0.5$), as inferred by Goldberg and Wo\'zniak (1998)   
from the OGLE data. Once the number $\bar n$ is given,
$\fo$ and $\fs$ are randomly determined for a given bulge luminosity function (LF).

For this latter, we use the LF obtained towards the Baade's Window with the Hubble Space Telescope (HST) (Holtzman et al. 1998).   
 Since there are very few observed bright stars, we have proceeded
as follows to obtain a combined LF: for bright stars ($V<18.5$) the HST LF is
matched to a power law function, with a $-2$ exponent.
We restrict our calculations to sources fainter than $V_{{\mathrm{sup}}}=16$, which corresponds to the
majority of observed sources (Alcock et al. 1997),   
 and we performed calculations
with different values of this parameter ($V_{{\mathrm{sup}}}$ ranging from $15.5$ to $16.5$).

The faint cutoff $V_{{\mathrm{inf}}}$ is more difficult to determine. We tried different limit
magnitudes between $V=21.5$ and $V=22.5$,
which correspond to the lower cutoff of the MACHO LF (Alcock et al. 1997).   
 Since we are only concerned by source
blending and not by amplification bias, there is no effect arising from the part of the LF below the detection limit.

\section{Effect on $\te$-histograms}

Blending affects the Einstein time histogram in two different ways. 
The first one is to decrease $\te$ to $\teobs$ (Eq \ref{Fteobs}).
The second one is to reduce the number of observed events:
if the maximum amplification is lower than the threshold amplification, the event is missed. This occurs if
 $B$ is lower than $B_{{\mathrm{min}}}$, where $B_{{\mathrm{min}}}$ is a function of $\umin$:
$$
B_{{\mathrm{min}}} \equiv \( \at - 1 \)  \( { \umintrue^2 + 2 \over \umintrue \sqrt{\umintrue^2+4} } -1 \)^{-1}
$$

The Einstein time histograms are calculated using a Monte-Carlo code originally developed by Chabrier and M\'era
 (M\'era, Chabrier \& Schaeffer 1998).  
  We have simulated $10^7$ events which result, once including detection efficiency, in about $\sim 10^6$ observed events.
The Galactic model (model $1$) used in the calculations is the following:
a bar-shaped bulge with a density
given by the model of Zhao (1996).    
The bar is in the plane of the disk, with an angle $\phi = 13^\circ$ with respect to
the Sun-Galactic center line, the near end of the bar lying in the first quadrant.
The bulge mass is taken to be $2.2 \times 10^{10} \msol$.

 The bar rotates rigidly with an angular speed $\Omega_{\rm bar}= 63.6\,  \mathrm{km/s/kpc}$, the velocity dispersion is
  $\sigma_\mathrm{bulge}=110\, {\mathrm{km}/\mathrm{s}}$ 
 in all directions.
The model of the disk  is a classical double exponential disk (Bahcall \& Soneira 1980),   
 with a scale length
$R_d = 2700$ pc and
a scale height $h=300$ pc. The local normalization of the disk is $\rho_\odot = 0.05 \,\msol/\mathrm{pc}^3$. The rotation velocity is 
$v_{\mathrm{rot,disk}}= \, 210 \mathrm{km}/\mathrm{s}$
and the isotropic velocity dispersion is $\sigma_{\mathrm{disk}}= 20 \,\mathrm{km/s}$.
The velocity distribution of each star is assumed to be gaussian, with the mean velocity equal to
the rotation velocity and the afore-mentioned dispersions.
The Sun is in the galactic plane, at a distance of $R_\odot=8$ kpc from the Galactic center. Its velocity is equal to the one of
the local standard of rest, $v_\odot = v_{\mathrm{LSR}} = 210 \,\mathrm{km/s}$. The visibility function of the sources is the one used
by Kiraga \& Pacsy\'nski (1994), 
 with $\beta = -1$.
The mass function is the one derived by M\'era et al. (1998).

We stress that the aim of the present paper is not to examine the validity of
different galactic models, but to study the effect of source blending on
the microlensing event time distribution and optical depth.

To compute the time-histograms, we also need to take into account the detection efficiency.
 We used the clump giant efficiency toward the bulge:
 this efficiency is equal to the sampling efficiency and is not corrected for blending 
  (Alcock et al. 1997).   
 
\begin{figure}
\psfig{file=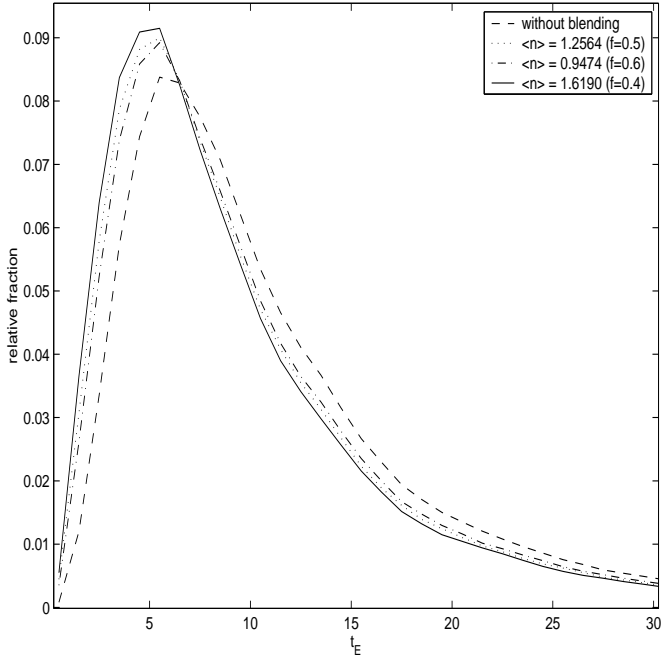,height=88mm,width=88mm}
\caption{Einstein time histogram for differents amounts of blending.
The curves give the relative number of short time events (below $66$ days), binned in one-day time intervals.
The curves are computed assuming a Poisson law with different values of $\bar n$ corresponding to different fractions of unblended stars ($\fb$).}
\label{effetB}
\end{figure}

Figure \ref{effetB} shows the effect of blending on the theoretical histogram 
(fraction of events as a function of the Einstein time) for different values of $\bar n$ in the Poisson distribution function of $B$ (\S 3).
 As expected, the effect of blending is to decrease the mean Einstein time when the fraction of unblended stars $\fb$  decreases. 
At the maximum of the histogram (between $4$ and $6$ days), there are $25 \%$ more events in the case of the Poisson DF ($\bar n \simeq 1.25$)
compared to the 
unblended histogram. This difference increases in the short-time region, and reaches $50 \%$ for $\te = 2.5$ days, but due to the decrease of the efficiency,
there are very few observed events in this region. The mean Einstein time is $10 \%$ to $15 \%$ lower
for the three Poisson DF.
If we convert this difference into a mean lens mass  ($\< \te \> \propto  \< \sqrt{m}\> $),
 we obtain a difference larger than $20 \%$ between the unblended case and
the Poisson law, {\it i.e.} the observed mean mass (derived from observations assuming that there is no blending) and the real mean mass
(that could be derived from observations if one knows the amount of blending)
i.e. $ \< m \>_{\rm obs} / \< m \>_{\rm real} < 0.8$.

\section{The observed optical depth}
\label{depth}

Source blending has three effects on the estimation of the optical depth $\tau$. The first one is to lower
$\te$, which tends to underestimate the optical depth. The second one, as we just showed, is to
underestimate the exposure (when a single source is thought to be observed, there are in fact
many sources that may be lensed), which tends to overestimate the optical depth. The last one is
due to the efficiency $\varepsilon$ which, depending on the variations of the efficiency with $\te$, can
either underestimate or overestimate $\tau$.
In order to quantify the resulting global effect, we performed the same Monte-Carlo calculations as described previously.

We first need to relate the true exposure to the observed one. If $n(i)$ is the real number of sources
inside the observed source $i$, the ratio of the true to the observed exposure is given by:
$$
{ E \over E_{{\mathrm {obs}}} } = { {\sum_{{\mathrm {sources}}}^{} } n(i) \over N_{{\mathrm {sources}}} }
$$
where $N_{{\mathrm {sources}}}$ is the number of monitored sources. The true exposure is given by
$$
E = \sum_{n=1}^{\infty} n \times N(n)
$$
where $N(n) = N_{{\mathrm {sources}}}\times  P(n) $ is the number of observed sources composed of $n$ blended sources. Therefore
$$
{ E \over E_{{\mathrm {obs}}} } =  \sum_{n \ge 1}^{} n P(n)  = { \sum_{n \ge 1}^{} n \wp(n) \over 1 - \wp(0) }  = {\bar n \over 1 - e^{- \bar n}}
$$

It is then straightforward to compute the ratio of the observed optical depth to the real optical depth,
for a given DF of the blending $B$. The luminosity of the source is
calculated for each event with the bulge LF. Then, knowing the real number $n(i)$ of sources, we can calculate the amount of blending.
 The ratio of the true to observed optical depth thus reads:
$$
{\tau \over \tau_{{\mathrm {obs}}}} =  { E_{{\mathrm {obs}}} \over E } {  \sum { \tetrue  \over \varepsilon(\tetrue) } \biggm/  
 \sum  { \teobs \over \varepsilon(\teobs  ) } }
$$

\begin{figure}
\psfig{file=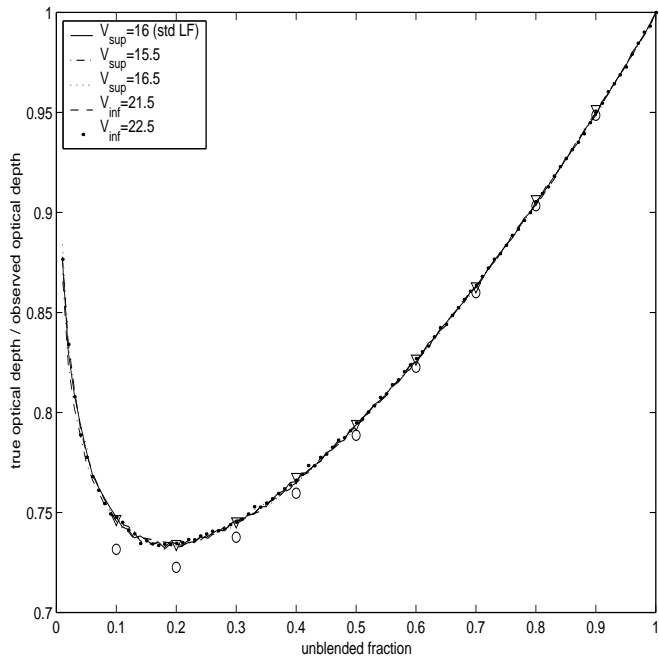,height=88mm,width=88mm}
\caption{Ratio $\tau / \tau_{\mathrm{obs}}$ as a function of the unblended source fraction for different values of the LF cutoff.
Standard values are $V_{\mathrm{inf}}=22$ and $V_{\mathrm{sup}}=16$ and these values are assumed except when indicated by the legend.
 Triangles and circles are computed using the same LF as the continuous line, but with model 2 and model 3 (see text). }
\label{tauB}
\end{figure}

Figure \ref{tauB} shows the ratio $\tau /\tau_{{\mathrm {obs}}}$ for different values of the fraction of unblended stars $\fb$. 
$\fb$ is related to the parameter of the Poisson law by
$$ \fb = { \bar n e^{- \bar n} \over 1 - e^{- \bar n} } = P(1) $$
In order to test the sensitivity of this ratio to the Galactic model, we have performed the same calculations
for two other models. Model 2 is obtained by changing only the bar angle (set to $\phi = 20 ^\circ$),
the mass of the bulge ($1.2 \times 10^{10} \msol$), and the disk scale length ($R_d =3500\, \mathrm{pc}$), while the local
normalization of the disk $\rho_\odot$ is unchanged.
Model 3 is obtained by changing the velocity parameters:
$(\sigma_{\mathrm{bulge}},\sigma_{\mathrm{disk}})=(100,30) \,\mathrm{km/s}$,
$\Omega_{\rm bar}= 50.0\,  \mathrm{km/s/kpc}$ and $v_{\mathrm{rot,disk}}=180\, \mathrm{km/s}$. As seen on Figure 2, 
 the ratio $\tau \over \tau_{{\mathrm {obs}}}$ is not very sensitive
to the Galactic model. This was expected since, forgeting the $\varepsilon$ term,
$\teobs = \te \times  g\(\umin,B\)$ with
$$
g \( \umin, B \) \equiv 
{ \sqrt{ f^2 \left({\at - \bbar    \over  B } \right) - \umintrue^2} \over \sqrt{ {\ut}^2 - {\uminobs}^2  }} \Theta \( B - B_{{\mathrm {min}}} \)
$$
where $\Theta$ is the Heaviside function (equal to $0$ for a negative argument).
Then the ratio $\tauobs / \tautrue$ can be written:
$$
{ \tauobs / \tautrue } =  {\bar n \over 1 - e^{- \bar n}} \< g \( \umin,B \) \>,
$$
\noindent which is independent of the galactic model.
The ratio is also quite insensitive to the cutoff values of the LF since all the curves in the case of model $1$ are very close to each other and
the differences are of the order of the errors in the Monte-Carlo calculations.

The effect of blending is to {\it overestimate the optical depth}, for each of the three models considered.
The ratio of the true optical depth to the observed one can reach $75\%$. This result is consistent with the one derived by Alard (1997)   
using a different mehod. By taking into account the blending directly in Eq. \ref{eqtau}, he found an overestimation of $\sim 15 \%$, whereas
we obtain $\sim 20 \%$ for $\fb = 0.5$.

We can relate our results to the last determinations of the optical depth using clump giant stars as sources.
Popowski et al. (2004) argue that, using clump giant and Differential Image Analysis, they can derive
an optical depth with no blending effect. This comes from the fact that, due to their luminosity, clump giant stars
are unlikely to suffer large amounts of blending. Note, however, that Sumi et al. (2005) suggest that
even clump giant stars can suffer some amount of blending, at least in the OGLE-II sample.
Using the last MACHO value as the true optical depth, the
corresponding observed optical depth for {\it blended} sources, assuming a Poisson law with $\fb \simeq 0.5$,
would be $\tau_{-6} = 2.71 ^{+0.59}_{-0.47}$, compatible with the value obtained from main sequence sources, $\tau_{-6} = 2.91 ^{+0.47}_{-0.45}$ (Alcock
et al. 2000).   

Note that for very crowded fields, $\bar n > 10$, the ratio $\tauobs / \tautrue$ is greater than one, so that in that case blending effects yield an {\it underestimation}
of the optical depth.

\section{Conclusion}

We have shown in this paper that blending can have an important effect both on the Einstein time histograms and
the inferred optical depth. 

For $\te$ histograms, the effect is to lower the observed $\te$ and then to decrease
the derived mean lens mass.
The change in the histogram is about $15 \%$ in the low-$\te$ region
corresponding to the maximum of the expected histogram, and can decrease substantially
the inferred mean lens mass (the difference can reach  $20 \%$).

For the optical depth, we have shown that, for all values considered for $\bar n$, the effect
is to overestimate the optical depth as a result of an underestimation of the exposure. 
In the case of the Poisson law we used, with an unblended fraction around $50 \%$, this effect can reach $20 \%$.
These results are quite insensitive to the cutoff of the LF and to the galactic model. We note, however, that in the case of very high blending,
for $\bar n > 10$, the effect of blending is to underestimate the optical depth. This can occur for microlensing experiments towards very crowded fields,
like M31 (AGAPE project, Ansari et al. 1997 
 and references therein).

Finally, to relate the observed optical depth to the one derived from different models, 
it is necessary to combine these results with the effect of other biases, like lens blending
(Han 1998)  
and amplification bias (Han 1997,   
 Alard 1997).  
The effect of lens blending is to underestimate the optical depth by about $10 \%$,
which can cancel, at least in part, the effect of source blending derived presently in the case of moderate blending (for the Poisson law used
in these calculations, this corresponds to an unblended fraction larger than $80\%$).
For amplification bias, Han (1997)  
 has shown that the effect would be to overestimate the optical depth. Therefore, the net effect of the various blending effects is very likely to overestimate the correct optical depth by a substantial fraction.
 Taking blending effects into account brings into agreement the optical depth
derived with main sequence sources (Alcock et al. 2000,   
  Udalski et al. 1994), the one derived using brighter stars
(Popowski et al. 2004, Sumi et al. 2005), and the one
derived
from usual Galactic models
(Bissantz et al. 1997,   
 Englmaier et al. 1999, Sevenster et al. 1999) at the 2$\sigma$ level, and thus reconciles
 experimental and theoretical determinations without drastic changes in galactic modelling. 
Further observational determinations of the exact
amount of blending are needed to nail down precisely
the net effect of blending.

\end{document}